\documentclass[11pt,twoside]{article}
\usepackage{asp2014}
\usepackage{color}

\aspSuppressVolSlug
\resetcounters

\newcommand{\magphys}{\textsc{magphys}}

\begin{document}

\title{(O4.4) Machine Learning for Scientific Discovery}

\author{Shraddha~Surana$^1$, Yogesh Wadadekar$^2$ and Divya Oberoi$^2$}
\affil{$^1$ThoughtWorks Pvt. Ltd., Binarius Building, Shastrinagar, Yerawada, Pune 411006, Maharashtra, India. \email{shraddha.surana@thoughtworks.com}}

\affil{$^2$National Centre for Radio Astrophysics, Tata Institute of Fundamental Research, Post Bag 3, Ganeshkhind, Pune 411007, India.}

\paperauthor{Shraddha~Surana}{shraddha.surana@thoughtworks.com}{0000-0002-3009-3178}{ThoughtWorks Pvt. Ltd.}{Engineering for Research}{Pune}{Maharashtra}{411006}{India}
\paperauthor{Sample~Author2}{Author2Email@email.edu}{ORCID_Or_Blank}{Author2 Institution}{Author2 Department}{City}{State/Province}{Postal Code}{Country}



  
\begin{abstract}
  
Machine Learning algorithms are good tools for both classification and prediction purposes. These algorithms can further be used for scientific discoveries from the enormous data being collected in our era. We present ways of discovering and understanding astronomical phenomena by applying machine learning algorithms to data collected with radio telescopes. We discuss the use of supervised machine learning algorithms to predict the free parameters of star formation histories and also better understand the relations between the different input and output parameters. We made use of Deep Learning to capture the non-linearity in the parameters. Our models are able to predict with low error rates and give the advantage of predicting in real time once the model has been trained. The other class of machine learning algorithms viz. unsupervised learning can prove to be very useful in finding patterns in the data. We explore how we use such unsupervised techniques on solar radio data to identify patterns and variations, and also link such findings to theories, which help to better understand the nature of the system being studied. We highlight the challenges faced in terms of data size, availability, features, processing ability and importantly, the interpretability of results. As our ability to capture and store data increases, increased use of machine learning to understand the underlying physics in the information captured seems inevitable.

\end{abstract}

\section{Introduction}

Machine learning (ML) algorithms learn the information contained in the data and use it for the purpose of prediction, classification and clustering. As astronomical datasets are growing exponentially in size, ML techniques are becoming increasingly useful for creating models which will enable astronomers to expedite the process of astronomical discovery.  For example, ML algorithms can be used to improve performance (in terms of time and processing capacity) - as we show in prediction of star formation properties in Section \ref{supervised}.  They can be used to generate inferences by analysing large amounts of data to uncover their patterns. The outcome of the ML algorithm can be compared with the current model and be used to improve the understanding of the current models i.e. if the ML predictions for a small number of data points form outliers from the trend then they may point to the need of further investigation, as typically ML algorithms are designed to generalise on all data points.

Section~\ref{supervised} briefly describes the outcome of the supervised deep learning approach to predict three star formation properties of galaxies viz. stellar mass, star formation rate (SFR) and dust luminosity using broadband flux measurements. Section~\ref{unsupervised} presents our exploration of the unsupervised technique to discover information present in solar radio images.

\section{Discovering physical models based on supervised machine learning} \label{supervised}

In this work, our goal is to model and replicate the behavior of a specific stellar population synthesis code viz. \magphys{} \citep{magphys}. The best fit parameters that characterise the star formation histories of galaxies are determined using these models. We implement a supervised machine learning technique - deep learning - to mimic the behavior of the \magphys{} model to predict three important star formation properties - stellar mass, star formation rate and dust luminosity. The data used are from the GAMA (Galaxy And Mass Assembly) \citep{Driver2011} survey. We selected all galaxies with  $0<z\leq0.5$ and stellar mass $\geq 10^9 M_{\odot}$ from the GAMA catalog. After applying filters to remove noise from the data, we had 76,455  galaxies which  form  our  final  sample set to train the deep learning model.
Estimating the three star-formation parameters using the \magphys{} code for 10,000 galaxies would take $\sim100,000$ minutes (about 10 minutes per galaxy). As telescope technology improves, the size of galaxy samples with multi-wavelength data is ever increasing. A traditional stellar population synthesis code scales linearly with sample size. On the other hand, the deep learning model training, takes 3 to 30 mins depending on the free parameter being modelled, with most of the time taken up by the training phase, which is a one time effort. Once modelled, the time taken to predict the free parameters for new galaxies is negligible. This represents a huge savings in time, with potentially larger savings for samples from future large area imaging surveys.  
This model can further be modified to also give the confidence level of each prediction. Galaxies for which the confidence level of prediction is very low, can be further investigated and run with the standard stellar population technique. This outcome can be investigated further and incorporated in the deep learning model, enriching the model as it encounters more and more data.
The error of the deep learning model is 0.0577, 0.1643 and 0.1143 (in their respective units) for the logarithm of stellar mass, star formation rate and dust luminosity respectively which is calculated using $$error = \sigma(y_{actual}-y_{predicted})$$
The comparison of predicted values against  values from \magphys{} is shown in Fig.~\ref{fig1}. More details on the prediction of star-formation properties is available in \citet{surana2020}.

\articlefigurethree{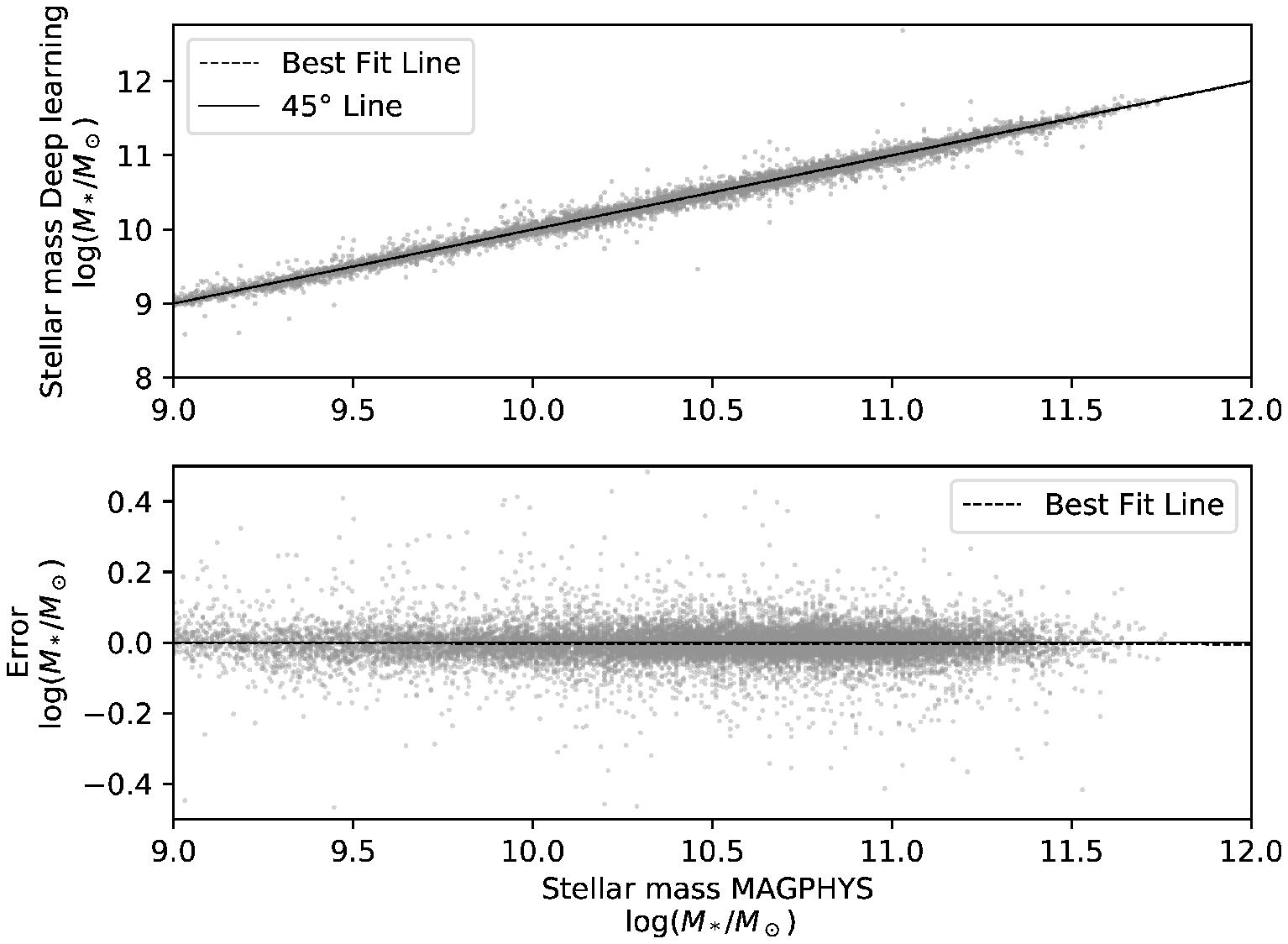}{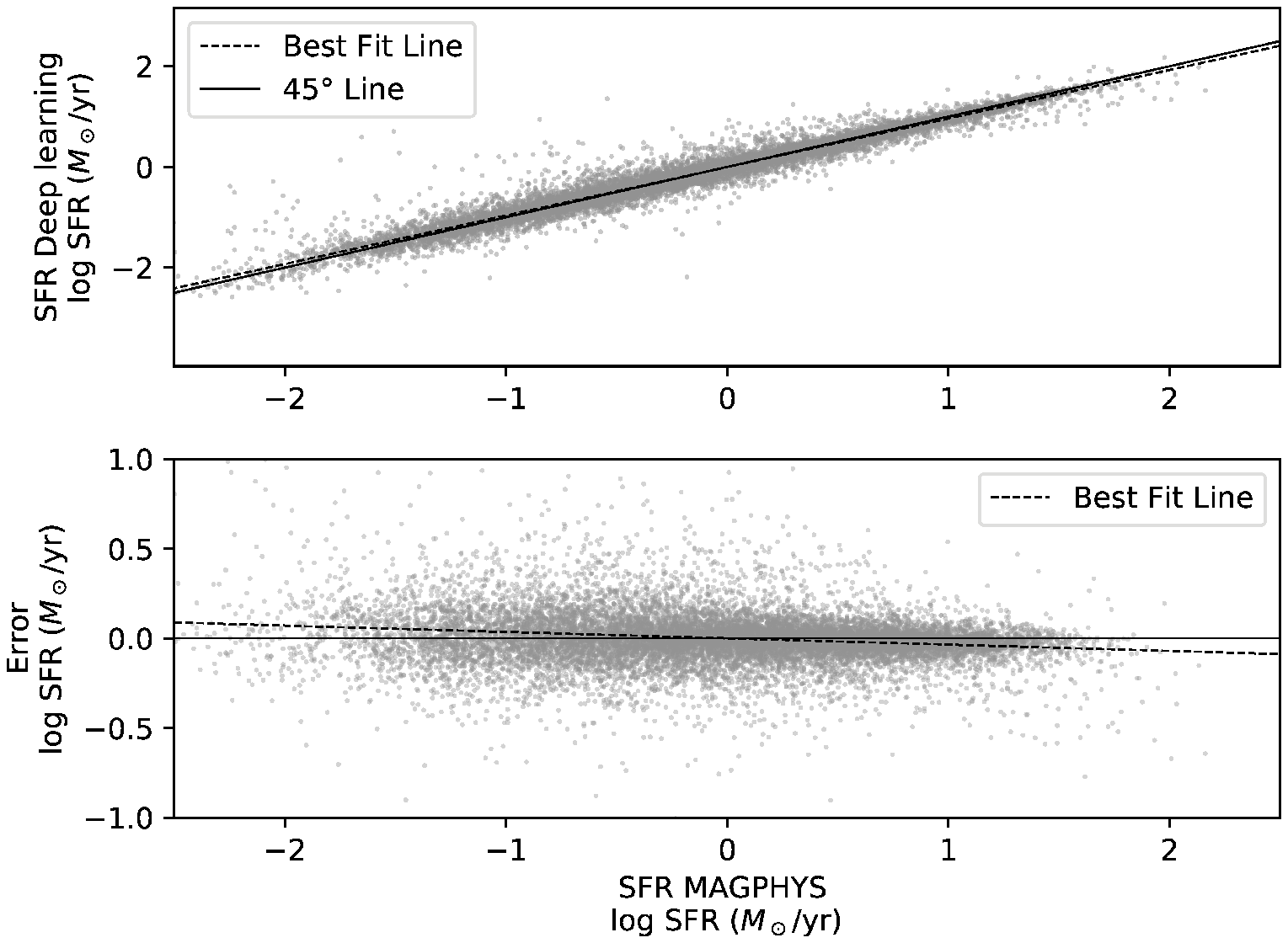}{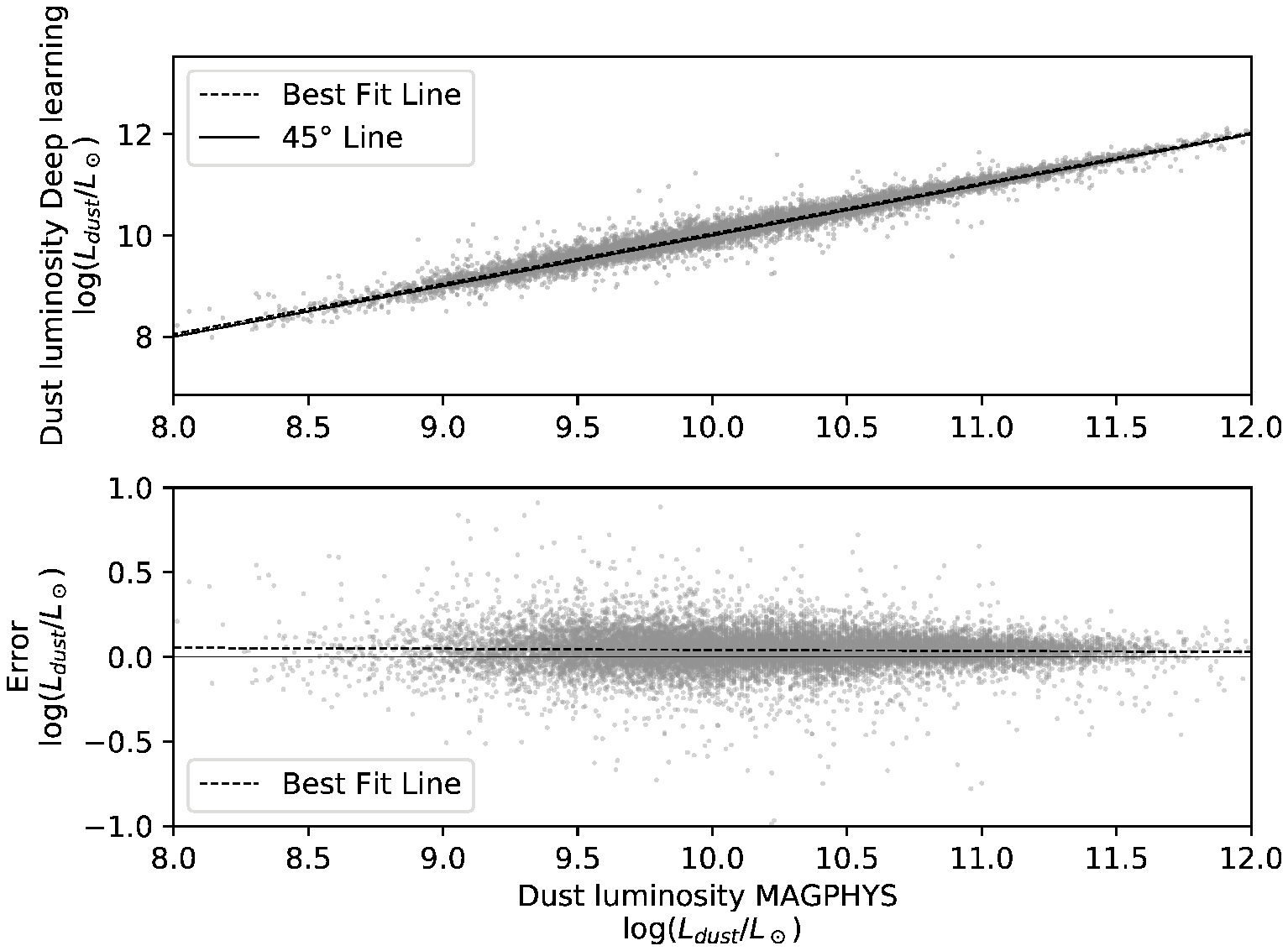}{fig1}{The scatter plots in the upper panels show the values predicted by the deep learning model compared to the \magphys{} values, for stellar mass, star formation rate and dust luminosity. The dashed line shows the best linear fit through the scatter plot and the solid line represents the points where the predicted values equal the \magphys{} model values. The lower panel shows the error in the prediction as a function of the respective parameter.}

\section{Discovering patterns and anomalies based on Unsupervised machine learning} \label{unsupervised}

Unsupervised ML algorithms are of particular importance in research,as they are used to uncover patterns, complex relationships and groupings that exist in the data. Quite often, the research explorations and approaches are biased by the knowledge, experience, expectations and intuitions of the researchers. Much of the analysis is also driven by visual inspection, which has severe limitations in terms of the number of dimensions that can be explored simultaneously, and the capacity of the human brain to assimilate information from large data volumes. We use unsupervised ML algorithms to understand data demography and observe groupings, patterns or anomalies in the data. We use solar radio images from the Murchison Widefield Array (MWA) to explore ML approaches. This opportunity has been enabled by the recent availability of an interferometric imaging pipeline \citep{Mondal-AIRCARS}, which provides solar radio images with unprecedented fidelity and dynamic range. The output of this pipeline is a hypercube - $I(\theta,\phi,\nu,t)$, where $I$ is the intensity of emission, $\theta$ and $\phi$ are the image coordinates, $\nu$ is the frequency, and $t$ is the time corresponding to the image. The MWA can produce up to $\sim 10^5$ images every minute. It quickly becomes infeasible to explore such large data sets using conventional human effort intensive approaches. Though, we are currently working with much slower image rate of $\sim 5.8\times 10^3$ images per minute, these are already challenging enough to require significant parallalization and processing power. This can be achieved both in the model implementation and the pre-processing of data required by the ML algorithm. In order to utilize this extensive amount of data and synthesise the information from them, we use an unsupervised ML technique called Self-Organising Maps (SOMs) \citep{kohonen-self-organized-formation-1982}. SOMs are a form of unsupervised neural networks that produce a lower dimensional (typically two) representation of the data, while preserving the topology. They can be used to explore similarities and/or dissimilarities in the solar images, pointing to the direction for further investigation. Visualization also forms a large part of the exploration, both before and after application of the unsupervised ML algorithms - before to understand the data and after to understand the patterns identified by the ML algorithm.

\section{Summary}
We have successfully applied Machine Learning to predict star formation properties of galaxies that has bought down the time to predict the specified parameters with very low error. Since ML models are designed to be generalised models, their results can be compared to existing models (simulations etc.) and the data points whose results do not match can be analysed further. This may help to better understand the underlying physical model, or help make the existing simulated models better. 
On the other hand, ML can be applied to a vast set of unstructured data to discover phenomena which would otherwise take a lot of time and insight to uncover. Our work in using SOMs in solar radio images, is a promising example in using ML techniques to uncover patterns, clusters and anomalies in the data. As our ability to capture and store data increases, increased use of machine learning to understand the underlying information captured is inevitable.

\acknowledgements SS thanks National Centre for Radio Astrophysics for hosting her for a part of this work. 
SS also thanks her colleagues from ThoughtWorks for the numerous interesting and useful discussions, and stimulating ideas. 

\bibliography{O4.4}

\end{document}